\documentclass[twocolumn,showpacs,preprintnumbers,amsmath,amssymb]{revtex4}

\usepackage{graphicx}
\usepackage{dcolumn}
\usepackage{bm}

\begin{document}
\title{Efficient routing on scale-free networks based on local information}
\author{Chuan-Yang Yin}
\author{Bing-Hong Wang}
\email{bhwang@ustc.edu.cn}
\author{Wen-Xu Wang}
\author{Tao Zhou}
\author{Hui-Jie Yang}
\affiliation{%
Department of Modern Physics and Nonlinear Science Center,
University of Science and Technology of China, Hefei, 230026, PR
China
}%

\date{\today}

\begin{abstract}
In this letter, we propose a new routing strategy with a single
free parameter $\alpha$ only based on local information of network
topology. In order to maximize the packets handling capacity of
underlying structure that can be measured by the critical point of
continuous phase transition from free flow to congestion, the
optimal value of $\alpha$ is sought out. By investigating the
distributions of queue length on each node in free state, we give
an explanation why the delivering capacity of the network can be
enhanced by choosing the optimal $\alpha$. Furthermore, dynamic
properties right after the critical point are also studied.
Interestingly, it is found that although the system enters the
congestion state, it still possesses partial delivering capability
which do not depend on $\alpha$. This phenomenon suggests that the
capacity of the network can be enhanced by increasing the
forwarding ability of small important nodes which bear severe
congestion.
\end{abstract}


\maketitle

Since the seminal work on the small-world phenomenon by Watts and
Strogatz \cite{WS} and scale-free networks by Barab\'asi and
Albert \cite{BA}, the evolution mechanism of the structure and the
dynamics on the networks have recently generated a lot of
interests among physics community \cite{Review1,Review2}. One of
the ultimate goals of the current studies on complex networks is
to understand and explain the workings of systems built upon
them\cite{Epidemic1,Epidemic2,Epidemic3,Cascade1,Cascade2,Yang},
and relatively, how the dynamics affect the network
topology\cite{WWX1,WWX2,WWX3,Zhu}. We focus on the traffic
dynamics upon complex networks, which can be applied everywhere,
especially the vehicle flow problem on networks of roads and the
information flow dynamic on interconnection computer networks.
Some previous works have focused on finding the optimal strategies
for searching target on the scale-free networks \cite{BJKim} and
others have investigated the dynamics of information flow with
respect to the packets handling capacity of the communication
networks \cite{loadKim,orderpara,optimal,
Tadic,Lai,alleviate,Yan}, however, few of which incorporate these
two parts. In this letter, we address a new routing strategy based
on the local information in order to both minimize the packets
delivering time and maximize the capacity of huge communication
networks.

In order to obtained the shortest path between any pair of nodes,
one has to know the whole network structure completely. However,
due to the huge size of the modern communication networks and
continuous growth and variance of the networks' structure, it is
usually an impossible task. Even though the network is invariant,
for the sake of routing packet along the shortest path each node
has to put all the shortest paths between any pair of nodes into
its routing table, which is also impractical for huge size because
of limited storage capacity. Therefore, In contrast to previous
works allowing the data packets forwarding along the shortest
path, in our model, we assume each node only has the topology
knowledge of it's neighbors. For simplicity, we treat all nodes as
both hosts and routers for generating and delivering packets. The
node capacity, that is the number of data packets a node can
forward to other nodes each time step, is also assumed to be a
constant for simplicity. In this letter, we set $C=10$.

\begin{figure}
\scalebox{0.80}[0.80]{\includegraphics{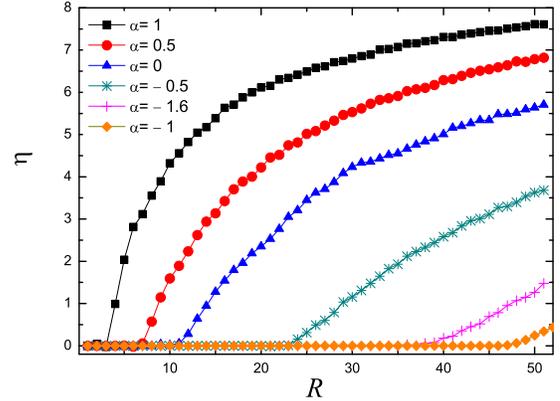}}
\caption{\label{fig:epsart} (color online). The order parameter
$\eta$ versus $R$ for BA network with different free parameter
$\alpha$.}
\end{figure}

Recent studies indicate that many communication networks such as
Internet and WWW are not homogeneous like random and regular
networks, but heterogeneous with degree distribution following the
power-law distribution $P(k)\thicksim k^{-\gamma}$. Barab\'asi and
Albert proposed a simple and famous model (BA for short) called
scale-free networks \cite{BA} of which the degree distribution are
in good accordance with real observation of communication
networks. Here we use BA model with $m=5$ and network size
$N=1000$ fixed for simulation. Our model is described as follows:
at each time step, there are $R$ packets generated in the system,
with randomly chosen sources and destinations, and all nodes can
deliver at most $C$ packets toward their destinations. To navigate
packets, each node performs a local search among its neighbors. If
the packet's destination is found within the searched area, it is
delivered directly to its target, otherwise, it is forwarded to a
neighbor node according to the preferential probability of each
node:
\begin{equation}
\Pi_{i}=\frac{k_i^\alpha}{\sum_jk_j^\alpha},
\end{equation}
where the sum runs over the neighbors of node $i$ and $\alpha$ is
an adjustable parameter. Once the packet arrives at its
destination, it will be removed from the system. We should also
note that the queue length of each node is assumed to be unlimited
and the FIFO (first in first out) discipline is applied at each
queue \cite{Lai}. Another important rule called path iteration
avoidance (PIA) is that a path between a pair of nodes can not be
visited more than twice by the same packet. Without this rule the
capacity of the network is very low due to many times of
unnecessary visiting of the same links by the same packets, which
does not exist in the real traffic systems.

\begin{figure}
\scalebox{0.80}[0.80]{\includegraphics{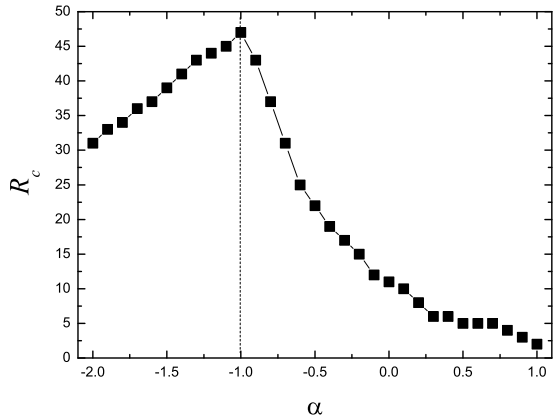}}
\caption{\label{fig:epsart} The critical $R_c$ versus $\alpha$.
The maximum of $R_c$ corresponds to $\alpha=-1$ marked by dot
line.}
\end{figure}

\begin{figure}
\scalebox{0.80}[0.80]{\includegraphics{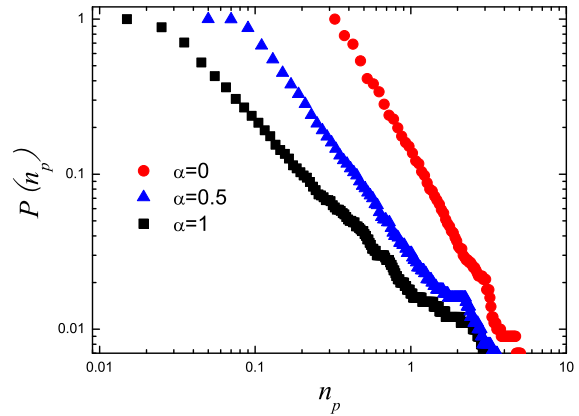}}
\caption{\label{fig:epsart} The queue length cumulative
distribution on each node by choosing different $\alpha$ more than
zero. Data are consistent with power-law behavior.}
\end{figure}

One of the most interesting properties of traffic system is the
packets handling and delivering capacity of the whole network. As
a remark, it is different between the capacity of network and
nodes. The capacity of each node is set to be constant, otherwise
the capacity of the entire network is measured by the critical
generating rate $R_c$ at which a continuous phase transition will
occur from free state to congestion. The free state refers to the
balance between created packets and removed packets at the same
time. While if the system enters the jam state, it means the
continuous packets accumulating in the system and finally few
packets can reach their destinations. In order to describe the
critical point accurately, we use the order
parameter\cite{orderpara}:
\begin{equation}
\eta(R)=\lim_{t\rightarrow \infty}\frac{C}{R}\frac{\langle\Delta
N_p\rangle}{\Delta t},
\end{equation}
where $\Delta N_p=N(t+\Delta t)-N(t)$ with $\langle\cdots\rangle$
indicates average over time windows of width $\Delta t$ and
$N_p(t)$ represents the number of data packets within the networks
at time $t$. For $R<R_c$, $\langle\Delta N\rangle=0$ and $\eta=0$,
indicating that the system is in the free state with no traffic
congestion. Otherwise for $R>R_c$, $\eta \rightarrow\infty$, the
system will collapse ultimately. As shown in Fig. 1, the order
parameter versus generating rate $R$ by choosing different value
of parameter $\alpha$ is reported. It is easy to find that the
capacity of the system is not the same with variance of $\alpha$,
thus, a natural question is addressed: what is the optimal value
of $\alpha$ for maximizing the network's capacity? Simulation
results demonstrate that the optimal performance of the system
corresponds to $\alpha\thickapprox -1$. Compared to previous work
by Kim et al. \cite{BJKim}, one of the best strategies is PRF
corresponding to our strategy with $\alpha=1$. By adopting this
strategy a packet can reach its target node most rapidly without
considering the capacity of the network. This result may be very
useful for search engine such as google, but for traffic systems
the factor of traffic jam can not be neglected. Actually, average
time of the packets spending on the network can also be reflected
by system capacity. It will indeed reduce the network's capacity
if packets spend much time before arriving at their destinations.
Therefore, choosing the optimal value of $\alpha=-1$ can not only
maximize the capacity of the system but also minimize the average
delivering time of packets in our model.

\begin{figure}
\scalebox{0.80}[0.80]{\includegraphics{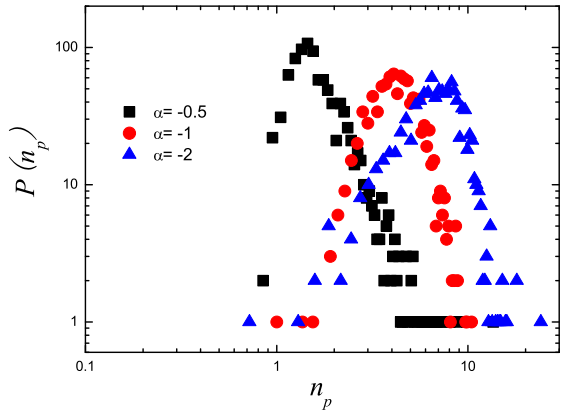}}
\caption{\label{fig:epsart} The queue length cumulative
distribution on each node by choosing different $\alpha$ less than
zero. $P(n_p)$ approximately exhibits a Poisson distribution.}
\end{figure}

\begin{figure}
\scalebox{0.80}[0.80]{\includegraphics{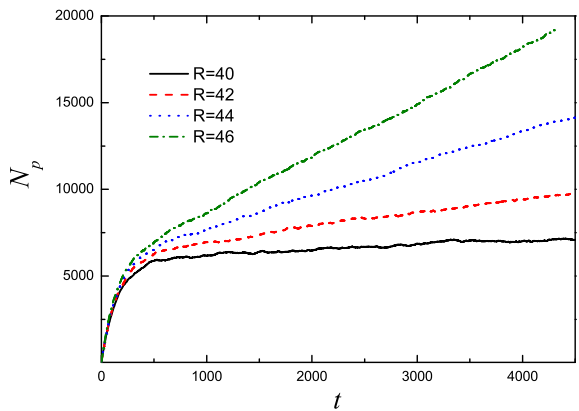}}
\caption{\label{fig:epsart} The evolution of $N_p$ for $R>R_c$.
Here, $\alpha_c$ takes $-1.5$ corresponding to the critical point
$R_c=39$.}
\end{figure}

To better understand why $\alpha=-1$ is the optimal choice, we
also investigate the distribution of queue length on each node
with different $\alpha$ in the stable state. Fig. 3 shows that
when $\alpha\geq 0$, the queue length of the network follows the
power-law distribution which indicates the highly heterogenous
traffic on each node. Some nodes with large degree bear severe
traffic congestion while few packets pass through the others and
the heterogenous behavior is more obviously correspondent to the
slope reduction with $\alpha$ increase from zero. But due to the
same delivering capacity of all nodes, this phenomenon will
undoubtedly do harm to the system because of the severe overburden
of small numbers of nodes. In contrast to Fig. 3, Fig. 4 shows
better condition of the networks with queue length approximately
displays the Poisson distribution which represents the homogenous
of each node like the degree distribution of random graph. From
this aspect, we find that the capacity of the system with
$\alpha<0$ is larger than that with $\alpha>0$. But it's not the
whole story, in fact, the system's capacity is not only determined
by the capacity of each node, but also by the actual path length
of each packet from its source to destination. Supposing that if
all packets bypass the large degree nodes, it will also cause the
inefficient routing for ignoring the important effect of hub nodes
on scale-free networks. By the competition of these two factors,
the nontrivial value $\alpha=-1$ is obtained with the maximal
network's capacity.

\begin{figure}
\scalebox{0.80}[0.80]{\includegraphics{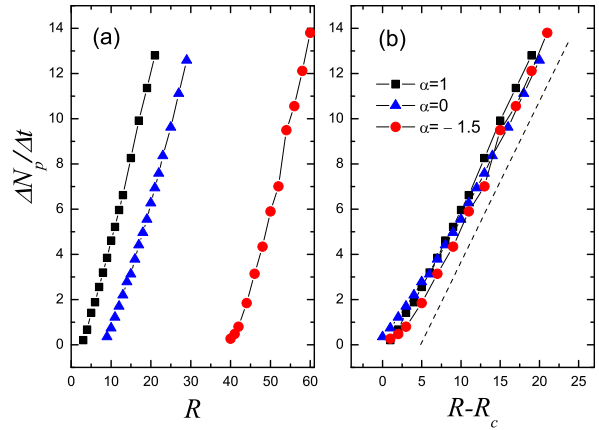}}
\caption{\label{fig:epsart} The ratio between $\Delta N_p$ and
time step interval $\Delta t$ versus $R$ (a) and versus $R-R_c$
the rescaling of $R$ (b) for different $\alpha$. In (b) three
curves collapse to a single line with the slope $\approx 0.7$
marked by a dashed line.}
\end{figure}

The behavior in jam state is also interesting for alleviating
traffic congestion. Fig. 5 displays the evolution of $N_p(t)$ i.e.
the number of packets within the network with distinct $R$.
$\alpha$ is fixed to be $-1.5$ and $R_c$ for $\alpha=-1.5$ is
$39$. All the curves in this figure can be approximately separated
into two ranges. The starting section shows the superposition of
all curves which can be explained by the fact that few packets
reach their destinations in a short time so that the increasing
velocity of $N_p$ is equal to $R$. Then after transient time,
$N_p$ turns to be a linear function of $t$. Contrary to our
intuition, the slope of each line is not $R-R_c$. We investigate
the increasing speed of $N_p$ with variance of $R$ by choosing
different parameter $\alpha$. In Fig. 6(a), in the congestion
state $N_p$ increases linearly with the increment of $R$.
Surprisingly, after $x$ axis is rescaled to be $R-R_c$, three
curves approximately collapse to a single line with the slope
$\approx0.7$ in Fig. 6(b). On one hand, this result indicates that
in the jam state and $R$ is not so large, the dynamics of the
system do not depend on $\alpha$. On the other hand the slope less
than $1$ reveals that not all the $R-R_c$ packets are accumulated
per step in the network, but about $30$ percent packets do not
pass through any congested nodes, thus they can reach their
destination without contribution to the network congestion. This
point also shows that when $R$ is not too large in the congestion
state, the congested nodes in the network only take the minority,
while most other nodes can still work. Therefore, the congestion
of the system can be alleviated just by enhancing the processing
capacity of a small number of heavily congested nodes.
Furthermore, we study the variance of critical point $R_c$
affected by the link density of BA network. As shown in Fig. 7,
increasing of $m$ obviously enhances the capacity of BA network
measured by $R_c$ due to the fact that with high link density,
packets can more easily find their target nodes.

\begin{figure}
\scalebox{0.80}[0.80]{\includegraphics{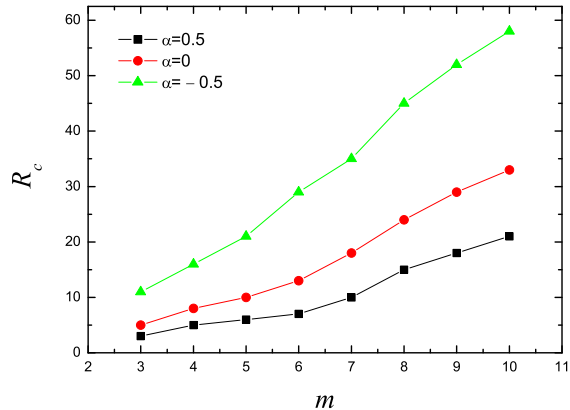}}
\caption{\label{fig:epsart} The variance of $R_c$ with the
increasing of $m$. }
\end{figure}

Motivated by the problem of traffic congestion in large
communication networks, we have introduced a new routing strategy
only based on local information. Influenced by two factors of each
node's capacity and navigation efficiency of packets, the optimal
parameter $\alpha=-1$ is obtained with maximizing the whole
system's capacity. Dynamic behavior such as increasing velocity of
$N_p$ in the jam state shows the universal properties which do not
depend on $\alpha$. In addition, the property that scale-free
network with occurrence of congestion still possesses partial
delivering ability suggests that only improving processing ability
of the minority of heavily congested nodes can obviously enhance
the capacity of the system. The variance of critical value $R_c$
with the increasing of $m$ is also discussed. Our study may be
useful for designing communication protocols for large scale-free
communication networks due to the local information the strategy
only based on and the simplicity for application. The results of
current work also shed some light on alleviating the congestion of
modern technological networks.

The authors wish to thank Na-Fang Chu, Gan Yan, Bo Hu and Yan-Bo
Xie for their valuable comments and suggestions. This work is
funded by NNSFC under Grants No. 10472116, 70271070 and 70471033,
and by the Specialized Research Fund for the Doctoral Program of
Higher Education (SRFDP No.20020358009).

\end{document}